# Application Of Support Vector Machines For Seismogram Analysis And Differentiation

## A DISSERTATION
*Submitted in partial fulfillment of the*

*requirements for the award of the degree*

*of*

**INTEGRATED MASTER OF TECHNOLOGY**
*in*
**GEOPHYSICAL TECHNOLOGY**

By
**Rohit Kumar Shrivastava**

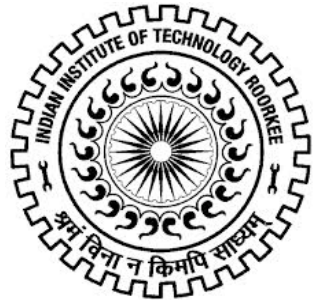

**DEPARTMENT OF EARTH SCIENCES
INDIAN INSTITUTE OF TECHNOLOGY ROORKEE
ROORKEE 247667 (INDIA)
JUNE, 2013**

# CANDIDATE'S DECLARATION

I hereby declare that the work, which is being presented in the dissertation entitled, '**APPLICATION OF SUPPORT VECTOR MACHINES FOR SEISMOGRAM ANALYSIS AND DIFFERENTIATION**" which is submitted as a partial fulfillment for the requirement of the award of the degree of **Integrated Master Of Technology** in **Geophysical Technology,** submitted in the **Department Of Earth Sciences, Indian Institute Of Technology, Roorkee** is an authentic record of my work carried under the guidance of Dr. Kamal, Professor, Department Of Earth Sciences, Indian Institute Of Technology, Roorkee and Dr. Roberto Carniel, Professor, Universita di Udine, Italy.

The matter embodied in the report has not been submitted to the best of my knowledge for the award of degree anywhere else.

Date:

Place : Roorkee                                                                               **Rohit Kumar Shrivastava**

This is to certify that the above statement made by the candidate is correct to the best of my knowledge.

Dr. Roberto Carniel                                                        Dr. Kamal
Professor                                                                          Associate Professor
Dipartimento Ingegneria                                              Department Of Earth Sciences
Civile e Architettura (DICA)                                        IIT Roorkee
Universita di Udine, Italy

# ACKNOWLEDGEMENT

I would like to thank **Dr. Kamal**, Associate Professor, Department Of Earth Sciences, Indian Institute Of Technology, Roorkee and **Dr. Roberto Carniel**, Professor, Department Of Civil Engineering And Architecture (DICA), University Of Udine, Italy, for their keen interest, constant guidance and encouragement throughout the course of this work.

I would also like to thank Nicola Alessandro Pino of INGV Naples, Italy for providing me with the seismic data.

Thanks are due to Dr. A.K Saraf, Professor and Head, Department Of Earth Sciences, Indian Institute Of Technology, Roorkee for providing various facilities during this dissertation.

I am also thankful to Dr. Mohammad Israil, Professor, Department Of Earth Sciences, Indian Institute Of Technology, Roorkee, (O.C. dissertation) for making the arrangement for the dissertation presentation and his support.

Last but not the least; I want to express my heartiest gratitude to my parents for their love and support which has been a constant source of inspiration.

Date :

Place : Roorkee

**Rohit Kumar Shrivastava**
**Integrated M.Tech – 5$^{th}$ year,**
**Geophysical Technology,**
**IIT Roorkee**

# ABSTRACT

Support Vector Machines (SVM) is a computational technique which has been used in various fields of sciences as a classifier with k-class classification capability, k being 2,3,4,…etc. Seismograms of volcanic tremors often contain noises which can prove harmful for correct interpretation. The PCAB station (located in the northern region of Panarea island, Italy) has been recording seismic signals from a pump installed nearby, corrupting the useful signals from Stromboli volcano. SVM with k=2 classification technique after optimization through grid search has been instrumental in identification and classification of the seismic signals coming from pump, reaching a score of 99.7149% of patterns which match the actual membership of class (determined through cross-validation) . The predicted labels of SVM has been used to estimate the pump's duration of activity leading to the declaration of corresponding seismograms redundant (not fit for processing and interpretation). However, when the same trained SVM was used to determine whether the seismogram used by Nicola Alessandro Pino et al, 2011 recorded at the same PCAB station on 4$^{th}$ April, 2003 contained pump's seismic signals or not, SVM showed 100% absence of pump's signals thereby authenticating the research work in Nicola Alessandro Pino et al.

# CONTENTS

# List Of Tables and Figures

## Tables :

## Figures :

# 1. Introduction

## 1.1 Motivation

Support Vector Machines (SVM) is a supervised classification technique used for classification of N types of data. SVM has been used in volcano seismology for determining different phases of volcanic eruption along with various other supervised and unsupervised classification techniques like Multi-layer Perceptron method , Cluster Analysis method, etc. where SVM stood out as the most efficacious tool. (Masotti et al, 2006; Langer et al, 2009; Cannata et al, 2011). During the training process, it recognizes the patterns, which is then used for classification. It is a non-probabilistic method, which can perform linear as well as non-linear classification through Kernel function transformation. Seismograms (graph generated by seismograph) contain various types of seismic signals from different types of ground motion. SVM can be used to classify and differentiate these signals, which can also prove to be efficient in identifying seismograms containing noises. Pump signals are often discrete in nature as compared to volcanic tremors (seismic signals generated by volcanoes) which are continuous in nature. SVM can be trained and tested to evaluate seismograms containing pump's seismic signals.

## 1.2 PCAB, Stromboli Volcano, Panarea Island, Italy.

The southern region of Tyrrhenian Sea has Aeolian islands of volcanic origin, located between Calabrian Arc region and oceanic basin's back arc. Volcanic seamounts and the archipelago form a horse-shoe shaped structure. These volcanoes are divided into three different areas [De Astis et al, 2003]: The western sector which has no activity repeated after 3040 ka, it includes the extreme western seamount, Filicudi and Alicudi islands ; the central sector which includes Volcano, Lipari and Salina islands having last eruption 1888-1890 A.D. , 580 A.D. and 13 ka b.p. , respectively ; the eastern sector which extends from the seamounts of the north eastern region to Panarea and Stromboli, it has fumarolic emissions which are intense in nature. Stromboli exhibhits active volcanism with occasional lava flows. Panarea's and surrounding islets's volcanic products date from 150 ka to 8 ka [Calanchi et al, 1999]. The subduction of Ionian region beneath Calabrian Arc lead to development of volcanism in these three different times, ranging from 0.8 Ma b.p. , 0.4 Ma to 1.3 Ma b.p. from east to west.

The rise of magma in these three sectors is a result of fault systems developed as a result of geodynamic regimes. The young magmas particularly of eastern sector comprising Panarea and Stomboli are enriched, metasomatised ithospheric mantle whereas, the western sector is poorly enriched asthenospheric. Apart from the similarities of the genesis of the magmas, the islands of eastern sector have displayed volcanic unrest in the past recent years. November 3, 2002 saw a considerable amount of $CO_2$-rich gas being discharged in the marine region surrounding Panarea.

PCAB station is a broadband seismic station installed in the northern region of Panarea by INGV Naples, Italy. The Stromboli island lies 20 km northwest to the station. PCAB seismic station has GURALP,CMG 40T-60s broadband seismometer.

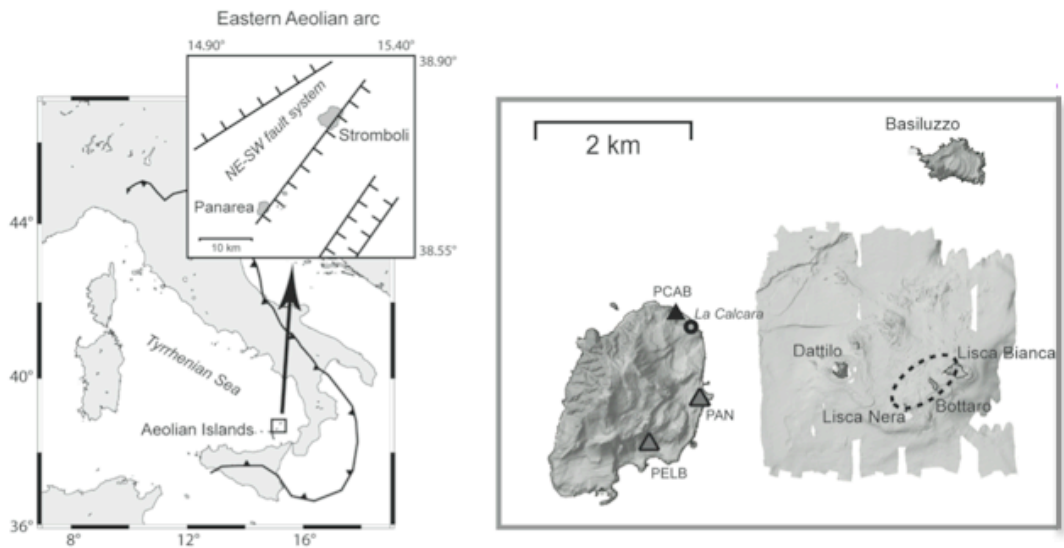

Figure 1: Map for the location of Panarea island and Stromboli island (Nicola Alessandro Pino et al, 2011)

**2**

## 1.3 Problem Statement

A pump is located at $38.638882^0$ North, $15.076461^0$ East at an elevation of 23m from the sea level which creates seismic signals being recorded at the PCAB station as noise, thereby rendering useful seismic signals from Stromboli volcano useless. The pump's signals being distinct and monotonic in nature have to be classified and isolated for useful interpretation of seismic signals from Stomboli volcano, Italy.

## 1.4 Proposed Model

Pump's duration of activity is identified for a particular day and the entire time series (seismogram) for the day is retrieved from the PCAB station . The seismogram containing the pump's seismic signals is sliced from the seismogram for the entire day which corresponds to the observed duration of activity of the seismogram. Another seismogram (of the same duration as the above sliced seismogram), which doesn't contain pump's seismic signals is sliced from the same seismogram for the entire day during which pump's duration of activity is known. Now using the spectrogram of these pump signals (seismogram containing pump's seismic signals) and pump-absent signals (seismogram not containing pump's seismic signals), Support Vector Machine is trained and optimized through Grid Search. Now the efficiency of this newly trained machine can be tested using Cross Validation (statistical technique). This machine is now ready for testing new data obtained from seismograms.

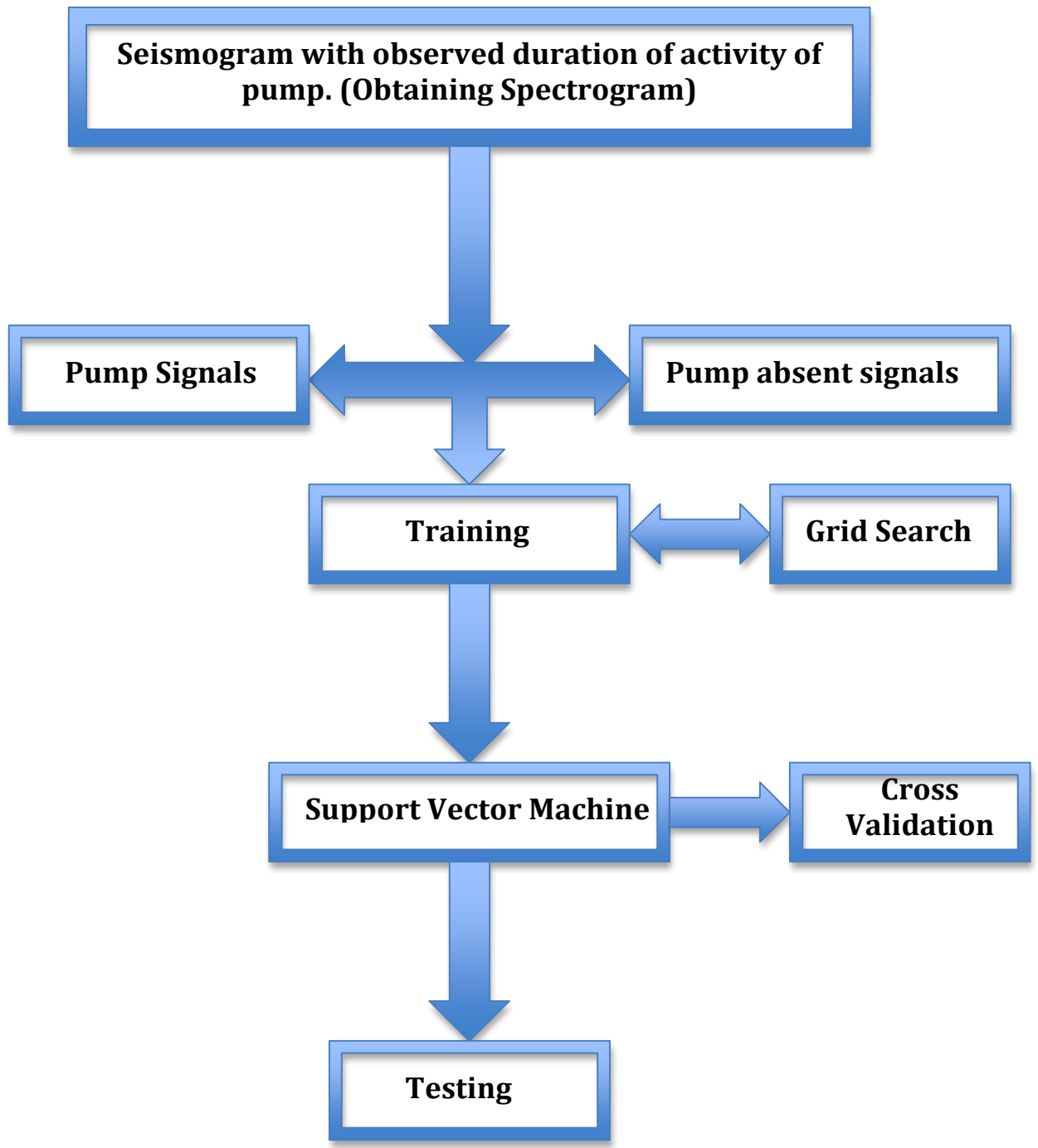

## 1.5 Software Packages Used

### Python

Python, a freeware software is a programming language used for scripting and developing application on various platforms in most of the areas. The interpreter and the library for many platforms are available in the binary or source form which is being freely distributed and can be downloaded from [www.python.org](www.python.org) .

Python interpreter can be easily extended to C++ or C with new data types and functions. For applications which can be customized, python can also be used as an extension language.

### ObsPy

ObsPy is a python tool. It is open source in nature and provides a framework in Python for studying and interpreting seismological data. There are various packages of ObsPy intended to serve their particular purpose. Eg. Obspy.core (for gluing the packages), obspy.imaging (for the purpose of imaging spectrograms, waveforms, etc.) various kinds of waveforms can be interpreted by ObsPy like css, datamark, gse2, mseed, sac, seisan, seg2, segy, sh and wav.

### Matlab

Matlab is a software used for technical computing, it has a very efficacious GUI and is being used in various fields of computational sciences. It is a licensed software and the matlab program used in this research work has been due to the license given to Indian Institute Of Technology, Roorkee for use by students. The Matlab eversion used is R2008a.

## Gnuplot

Gnuplot is a graphing tool for Mac OSX, Microsoft Windows, Linux, etc. driven by command line. Basically it was developed to simulate mathematical functions but now it has grown into various other dimensions such as using it as an engine for plotting by programs like Octave, Python, etc. Gnuplot is being developed since 1986.

# 2. Theoretical Aspects

## 2.1 Spectrograms

Spectrograms are tools of watching the variation of seismogram's frequency content with time. The seismogram's frequency spectrum is calculated between 0 to 31.25 Hz (as the sampling rate is 62.5 Hz) for the specified number of points. The spectral amplitude is represented by color varying from deep blue which represents low values, green, blue and then finally deep red which represents high values. After the frequency spectrum has been calculated for the specified number of points it is plotted as a vertical line. Now, these vertical line are plotted adjacent to each other with the specified amount of overlap (given by you) generating the frequency's spectrum time sequence.

Format : The most common one is a graph with 2 geometric dimensions: x-axis represents time, y-axis represents frequency, a $3^{rd}$ dimension which indicates the amplitude of frequency for a particular time, represented by the color of the point. Sometimes rather than using color to represent the amplitude, the spectrum can be plotted on a 3-dimensional surface where the height represents the amplitude of frequency for a particular time.

Spectrograms are plotted in two ways : a filterblank resulting from sequential bandpass filters (this was used before advent of digital signal processing), or calculated from Short Time Fourier Transform (STFT).

The spectrogram() function used in matlab :

Calling [y,f,t,p] = spectrogram(double(**data**),**window**,**overlap**,**fft**,**Fs**) generates the parameters needed for plotting a spectrogram of a seismogram in matlab, here **data** is the seismogram matrix, **window** is the space where an integer is specified, which is the number of segments in which double(data) is divided. For this segment-division Hamming window is used. **overlap** is used for overlapping segments before plotting. "**fft**" is used to specify the number of sampling points for carrying out discrete fourier trandform. "**Fs**" is used to specify the sampling frequency of the seismogram.

## 2.2 Pattern Recognition

If an input value (given) is assigned to a label in machine learning, it is known as pattern recognition. Classification is an example of Pattern Recognition. It attempts to assign an input value to a set of classes (given). Other examples of pattern recognition are regression, sequence labeling and parsing. Pattern Recognition is categorized into three types :

1. Supervised Learning
2. Semi-supervised Learning
3. Unsupervised Learning.

Supervised Learning uses Training Set (already provided), which consists of instances that have been correctly labeled by hand with the output. Unsupervised Learning identifies the pattern in the data, which is inherent in nature and has not been labeled preemptively. Semi-supervised Learning is using the combination of the above two methods.

**Support Vector Machine** uses **Supervised Classification** technique.

# 2.3 Supervised Classification: Support Vector Machines (SVM)

SVM algorithm was developed by Vapnik in 1995 and later introduced in soft computing in 1998. It's application is related to 2 concepts :

(i) For binary classification SVM uses lines, hyperplanes or even planes for differentiation. The function which differentiates the classes depends on the boundary points, the so called support vectors. Due to this property, SVM stands out amongst general classifier algorithms where the function differentiating the classes is defined by the distance of the centroids of the classes and the summed variance or co-variance of the classes. The hyperplane generated by SVM is at maximum distance from the classes defined by the training set. The margin which which separates the classes is of maximum width.

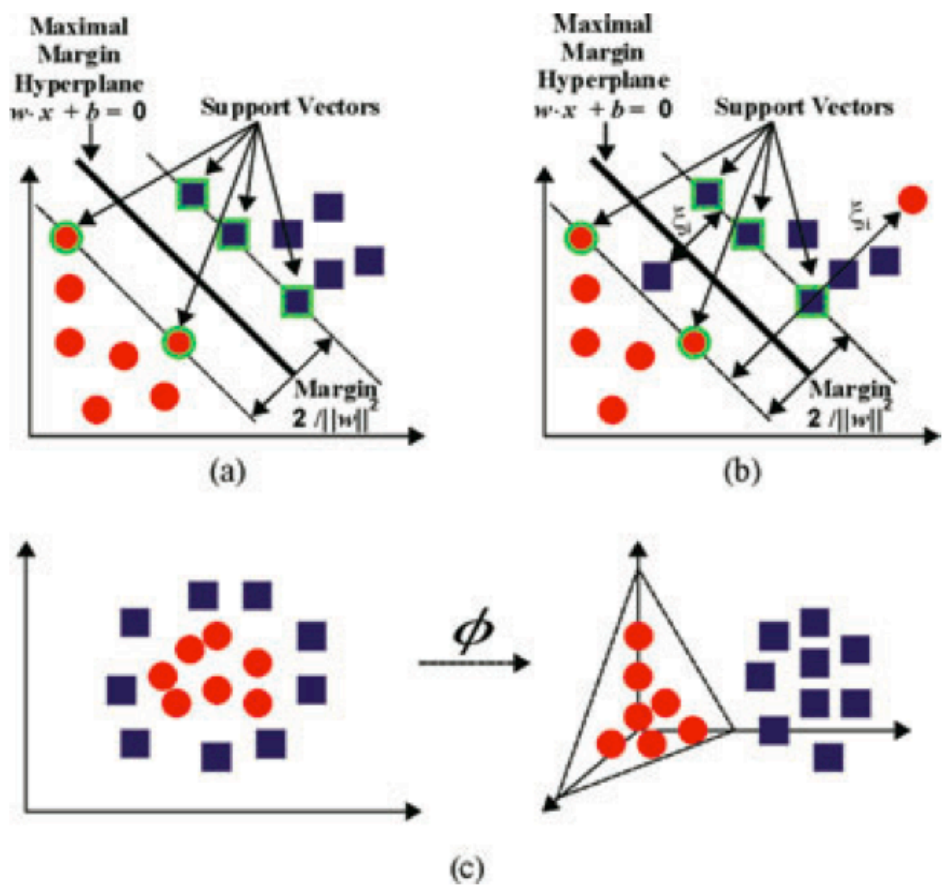

Figure 2 : Support Vector Programming (Langer et al,2009) : (a) linear classification (b) non-linear classification (c) Kernel function transformation

The maximum marginal hyperplane is defined by the function :

$$f(\mathbf{x}) = \text{sgn}(\mathbf{w} \cdot \mathbf{x} + b)$$

where $\mathbf{x} = (x_1, x_2, x_3, x_4, \ldots x_N)$ and N is a pattern present in the feature space. SVM assigns $f(x) = \pm 1$ label to the above mentioned pattern. SVM training basically deals with finding the weights $\mathbf{w} = (w_1, w_2, w_3, w_4, \ldots w_N)$ and b (bias), resulting in correct classification of N patterns while $2/||w||^{2}$ (margin between the classes) is maximum. The problem which has to be solved is as follows :

minimizing $||w||^2$ by keeping $y_i(x_i.w + b) \geq 1$ condition satisfied where i = 1,2,3,4,.N and $y_i = \pm 1$ represent the a priori classification of $x_i$.

$$L_p = ||\mathbf{w}||^2/2 - \sum_i \alpha_i \cdot (y_i \cdot (x_i \mathbf{w} + b) - 1)$$

By minimizing $L_p$ (Langrangian primal) the quadratic problem's solution can be obtained. Since the SVM generated function used for differentiation between classes in a solution of quadratic problem, the function generated by SVM is unparalleled. Misclassification might occur for few patterns, which can be accounted by using $\xi_i$ (error slack) and C (regularization parameter). The quadratic problem becomes :

Minimizing $||w||^2$ - $C\sum_i \xi_i$ while $y_i.(x_i.w+b) \geq 1- \xi_i$. Hence, after calculations :

$$f(\mathbf{x}) = \text{sgn}\left(\sum_j y_j(\alpha_j x_j \mathbf{x} + b) - 1\right)$$

$j \in (1,M)$, M is the number of support vectors.

(ii) For non-linear classification (figure 2(c)), a transformation is needed which maps the data to a feature space where SVM can easily carry out the classification process. The transformation is done by Kernel functions. The mapping is done by replacing the dot products $\mathbf{x_j}.\mathbf{x}$ by Kernel functions $\phi(x_j)$. $\phi(x) = K(x_j.x)$

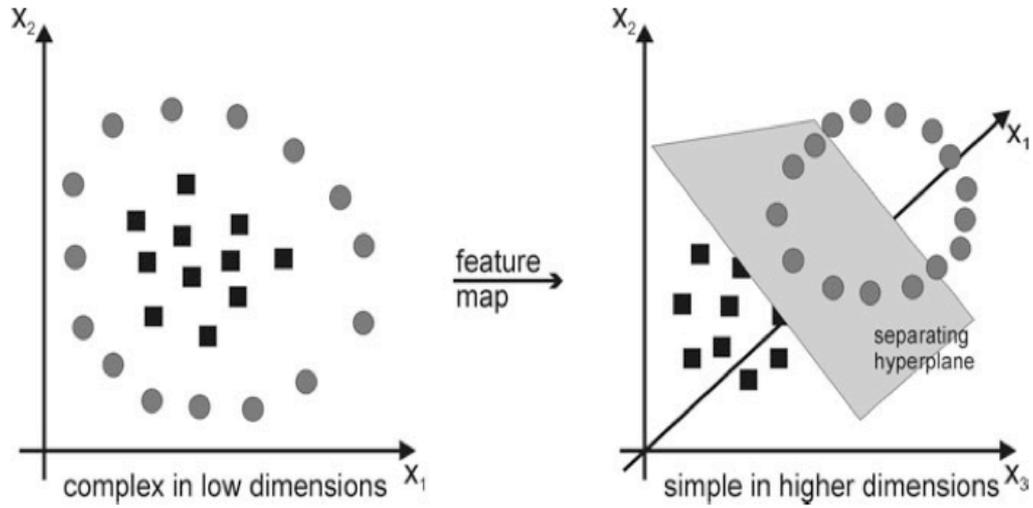

Figure 3 Kernel Transformation,(Cannata et al, 2011)

Some of the Kernel functions are as follows :

1) Linear function : $\mathbf{x_j.x}$
2) Polynomial function : $(\gamma x_j.x + r)^d$
3) (RBF) Radial Basis Function : $\exp(-\gamma\|x_j-x\|^2)$
4) Sigmoid function : $\tanh(\gamma x_j.x + r)$

## 2.4 Cross Validation

Cross validation also known as rotation estimation is a statistical tool to compare and evaluate learning algorithms. It divides the data into 2 segments, one which is used to train the model and the other which is used for validation. The basic method is k fold cross validation. The k fold cross validation has data partitioned into equally k sized folds or segments. Validation and training is done k times. In a single iteration, the machine the machine learns from k-1 folds and the remnant data or fold is used for validation. The learning algorithm's for each fold is measured in terms of accuracy and then after completion, the accuracy is averaged of all the k-iterations.

Cross validation strives to achieve :

1) The learned model's estimated performance.
2) Comparison of performance between 2 or more different algorithms.

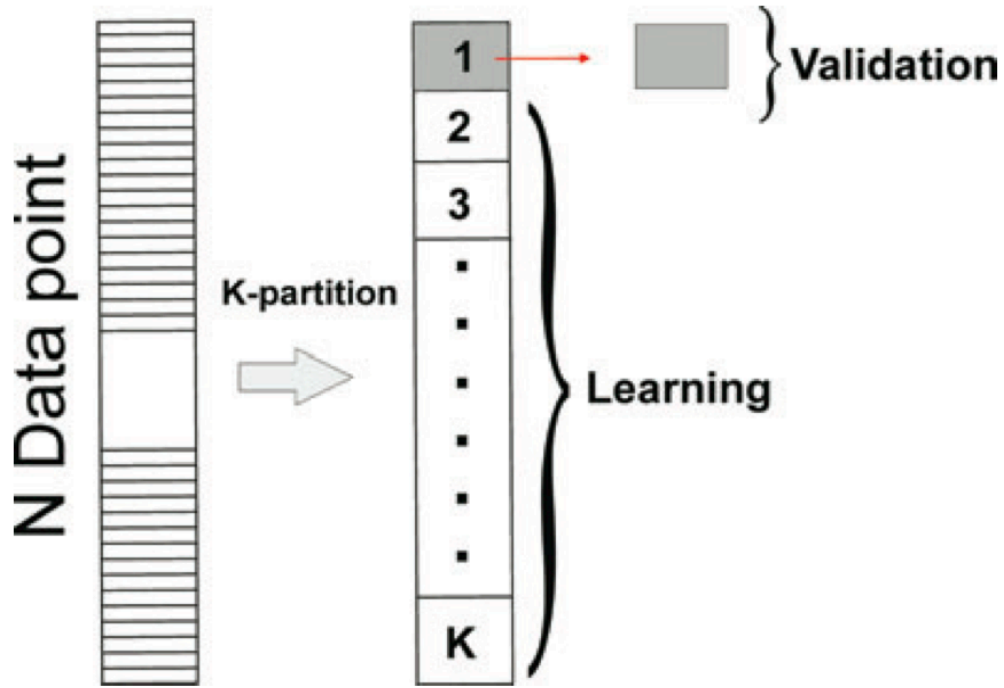

Figure 4 : Cross Validation theory, (Cannata et al,2011)

## 2.5 Grid Search

Grid Search is a form of hyperparameter optimization. It helps in selecting hyperparameters which increase the efficiency of a learning machine/algorithm. Hyperparameter optimization such as grid search ensures the trained model does not overfit the data (Regularization). Grid Search carries out an exhaustive search in the hyperparameter's space as specified by the training model or the learning algorithm. Grid Search is generally guided by cross validation which is carried out on training data. The R.B.F Kernel used over here by SVM classifier has two parameters C and γ which need to be evaluated by grid search. However, before using grid search one needs to discretize and set the boundaries of the parameters manually and the continuous search is performed.

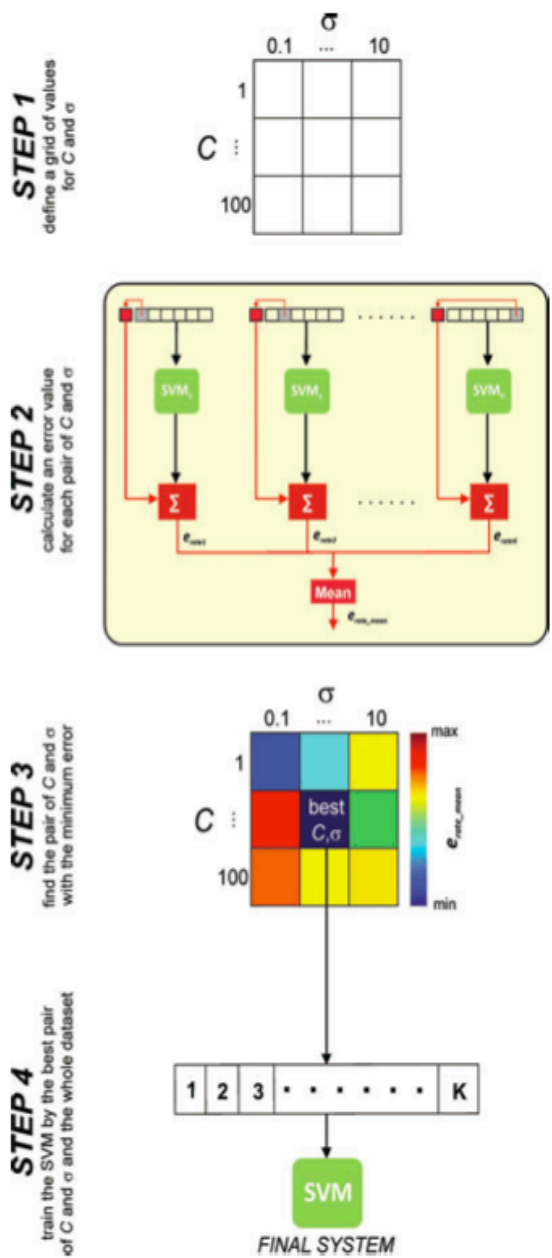

Figure 5 : Grid Search ( Cannata et al,2011)

# 3. Applications And Results.

## 3.1 Seismic Data

PCAB data has been installed by INGV Naples , Italy in the northern region of Panarea island which is a Guralp , CMG 40T-60s broadband seismometer (Figure 1) . The station has been the source of seismic data used for training by SVM classification technique . On $5^{th}$ December 2002 , the pump was observed to be switched on from 06:55 to 10:02 (Universal Time Co-ordinate , UTC) which has been used for the purpose of training (Figure 2) . Now , for testing the same time series (for the entire day) has been used which was used for training . Another time series of seismic signals recorded at PCAB station on $7^{th}$ December 2002 which had pump signals switched on at unknown intervals has been tested by the previously mentioned trained machine . The volcano seismic signal used by Nicola Alessandro Pino et al , 2011 recorded at the same seismic station on $4^{th}$ April , 2003 where the seismic precursors were used for tracking deep gas accumulation and slug upraise has also been tested by the machine.

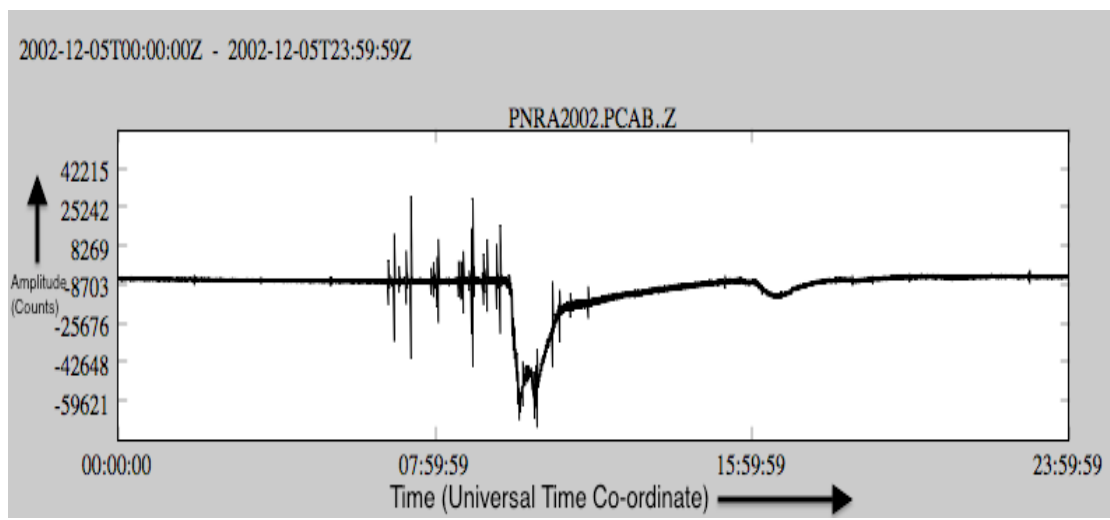

Figure 6 : Seismogram of 5th December 2002

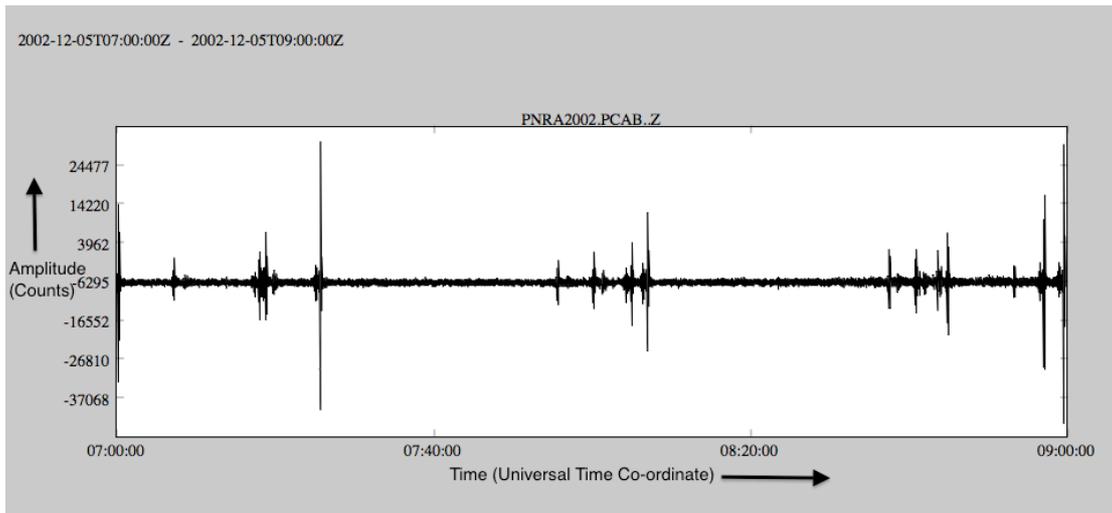

**Figure 7 : pump's seismogram**

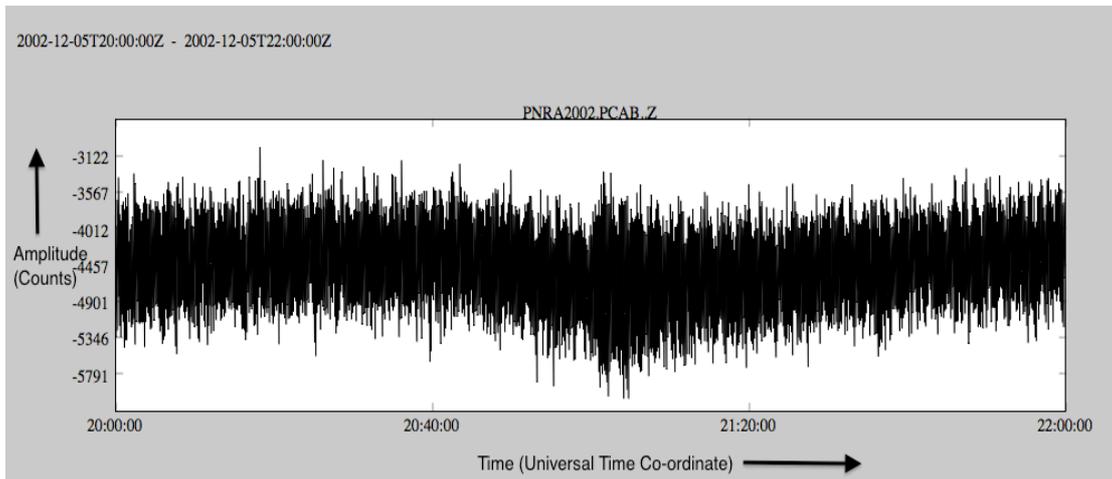

**Figure 8 : seismogram absent pump**

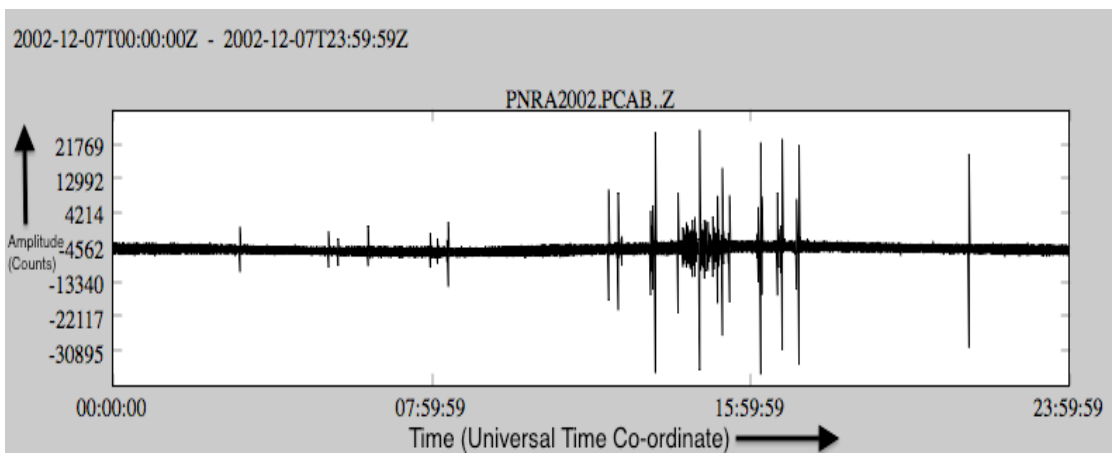

**Figure 9: seismogram of 7th December , 2002 with diverse interruption of pump's signals**

**15**

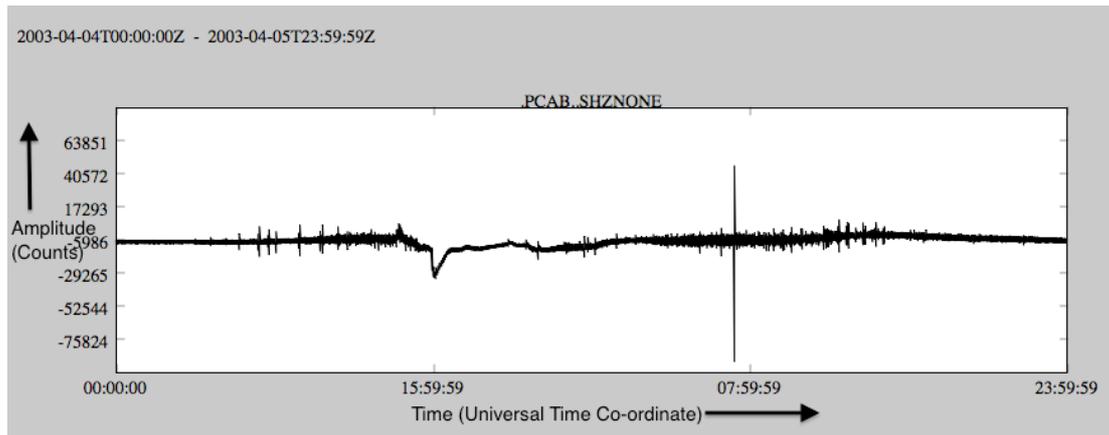

Figure 10 : Seismogram used by Nicola Alessandro Pino et al, 2011

## 3.2 Training

For training the SVM, Power Spectral Density feature vectors of seismogram recorded on 5[th] December 2002 from 07:00 to 09:00 (UTC) (Figure 7) containing pump signals and 12:00 to 14:00 (UTC) (Figure 8) which do not contain pump signals have been used. The pump signal containing seismogram ( 07:00 to 09:00 ,UTC) has been labeled as +1 and the other (12:00 to 14:00 , UTC) has been labeled as -1 when the pump was observed to be switched on and off respectively .The Kernel function used here is Radial Basis Function whose C and $\gamma$ parameters have been calculated by grid search (an optimizer technique). C = 2048.0 and $\gamma$ = 0.0001220703125 optimizes the machine giving 99.7149% correct classification of the parent class determined through cross validation. SVM1 is the trained model for testing data with a sampling frequency of 62.5 Hz, which is the sampling frequency of the data used for training set. SVM2 is the trained model for testing data with a sampling frequency, 31.25 Hz (seismogram used by Pino et al, 2011) , the sampling frequency of training set data is reduced to 31.25 Hz and then used for training the model. The SVM applied here uses 200 X 929 points of frequency – time dimensional vectors for training . For getting homogenious feature vectors we normalized (averaged) the rows of  200 X 929 dimensional matrix.

### 3.2.1 Grid Search

A hyperparameter optimization technique i.e. Grid Search was carried out to determine the most efficacious C and $\gamma$ parameters .

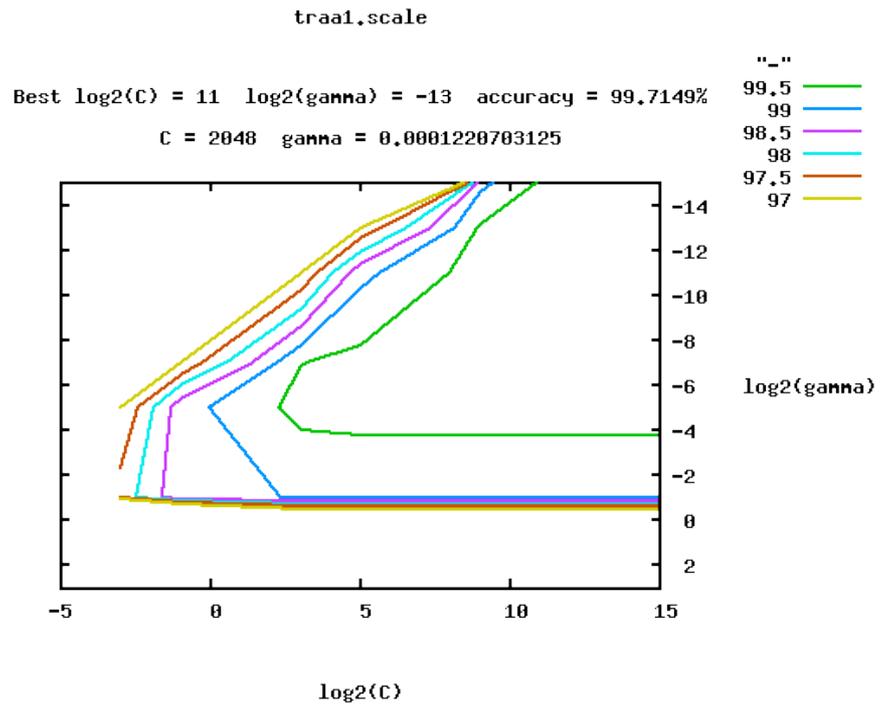

**Figure 11: Grid search for SVM1**

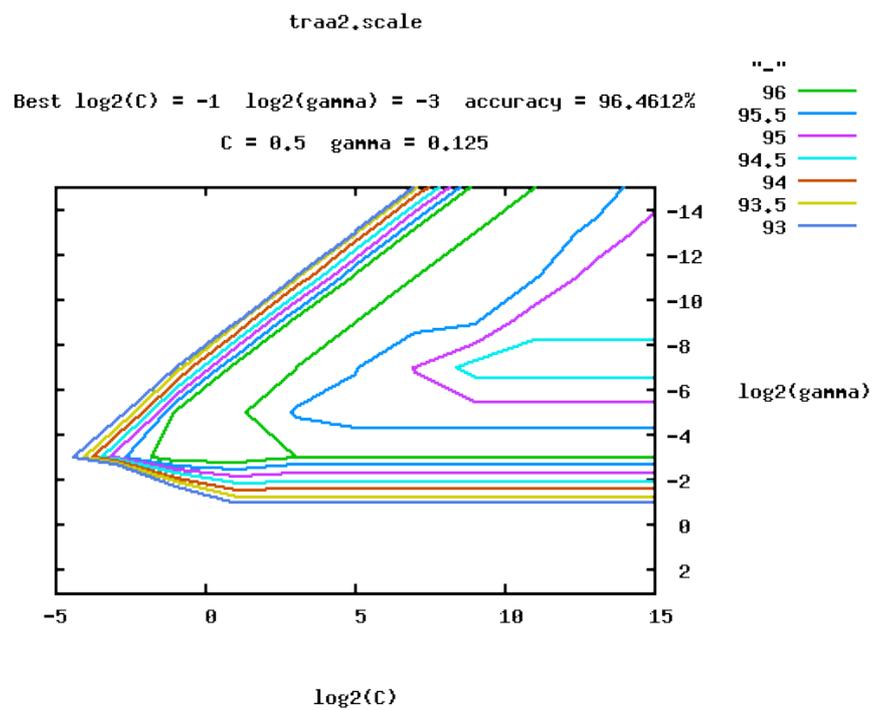

**Figure 12 : Grid Search for SVM2**

### 3.2.2 Cross Validation

Cross Validation of SVM1 and SVM2 has been carried out along with grid search. The cross validation performed over here is a k fold cross validation with k = 5.

**SVM1 :**

Table 1 : Grid Search and Cross Validation for SVM1

| C (Regularization parameter) | γ (Kernel Parameter) | Accuracy (5 fold Cross Validation) |
|---|---|---|
| 32.0 | 0.078125 | 99.6579% |
| 2048.0 | 0.0001220703125 | 99.7149% |

**SVM2 :**

Table 2: Grid Search and Cross Validation for SVM2

| C (Regularization parameter) | γ (Kernel Parameter) | Accuracy (5 fold Cross Validation) |
|---|---|---|
| 32.0 | 0.078125 | 95.5479% |
| 2048 | 0.0001220703125 | 95.6621% |
| 0.5 | 0.125 | 96.4612% |

## SVM1 Training

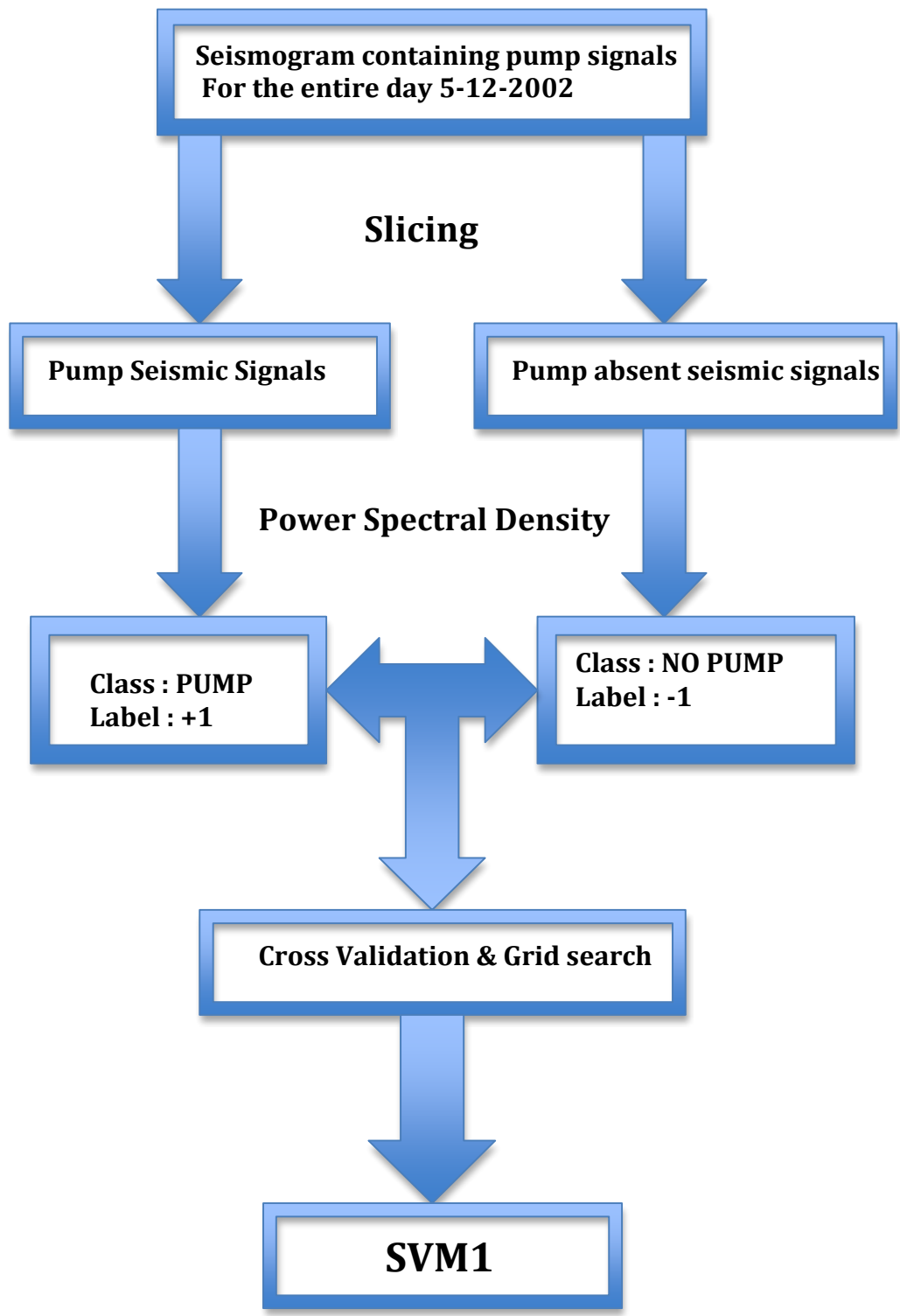

**SVM2**

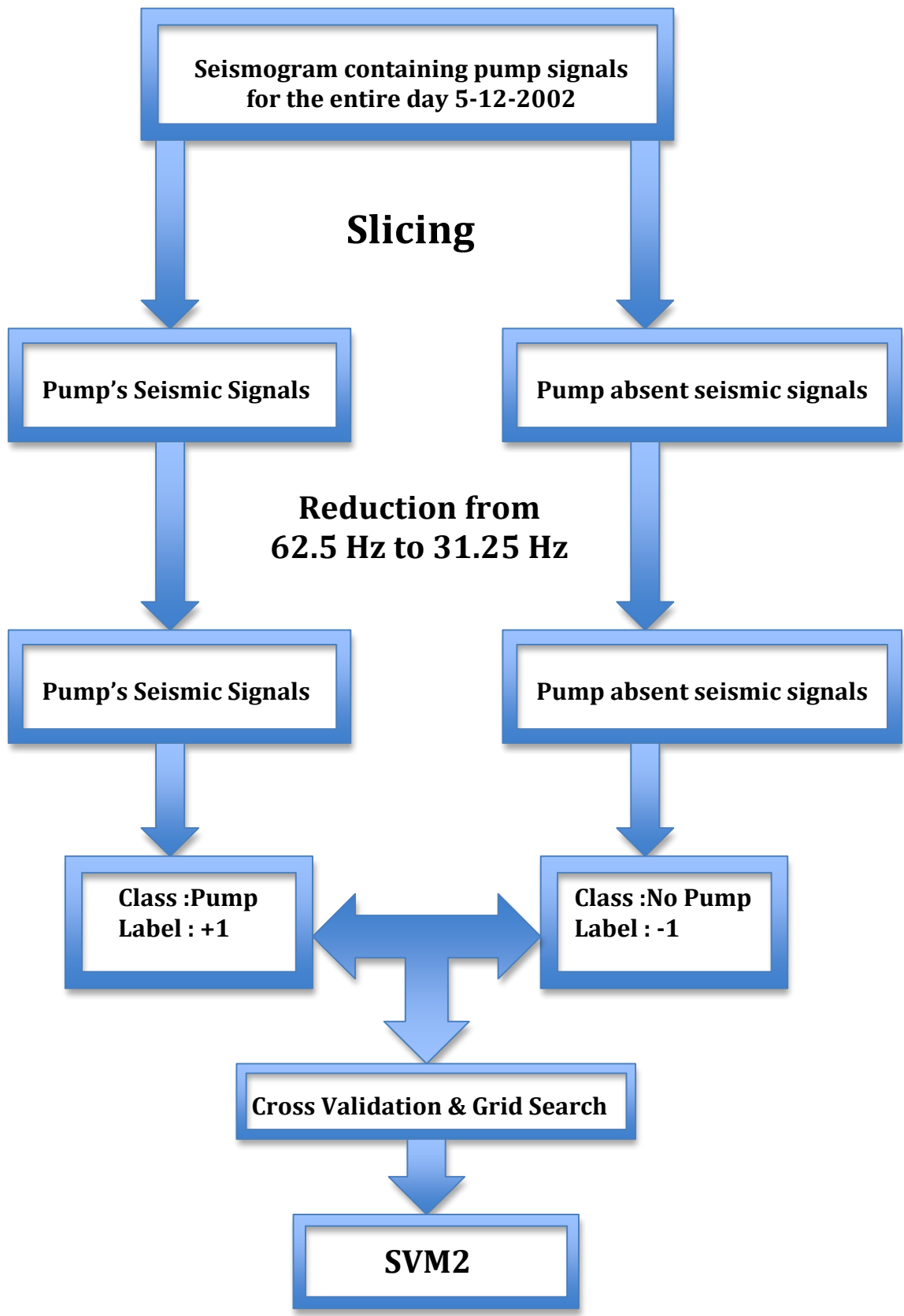

## 3.4 Testing

SVM1 and SVM2 can be used on spectrograms of seismograms for testing the presence of pump's seismic signals. SVM1 machine is used on seismograms with sampling frequency of 62.5 Hz and SVM2 machine is used on seismograms with sampling frequency, 31.25 Hz.

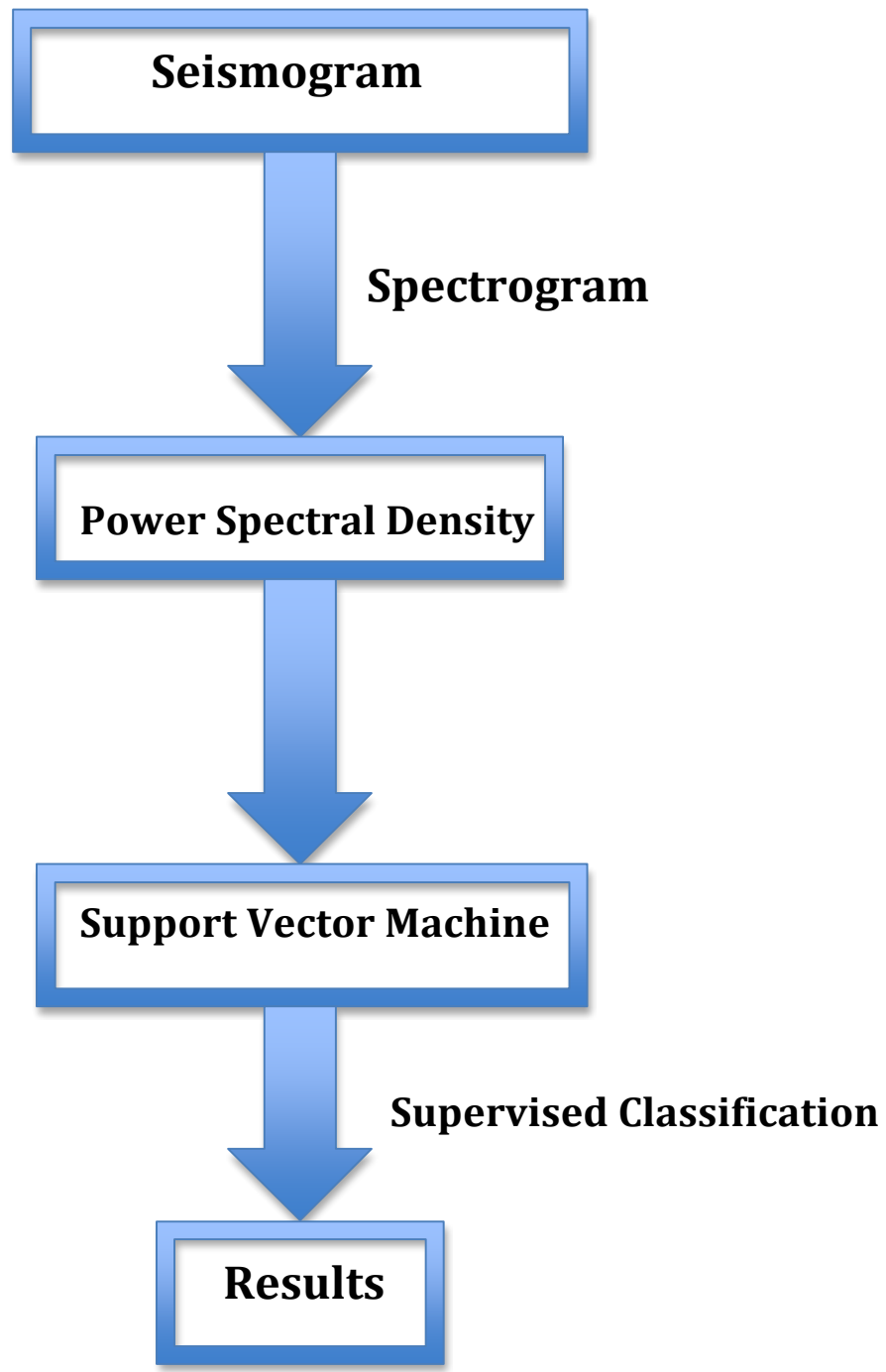

### 3.4.1 Testing of original data

Testing the seismogram for the entire day recorded on 5$^{th}$ December, 2002 by SVM1 gives a result that 92.557% of the signals do not contain pump's seismic signals.

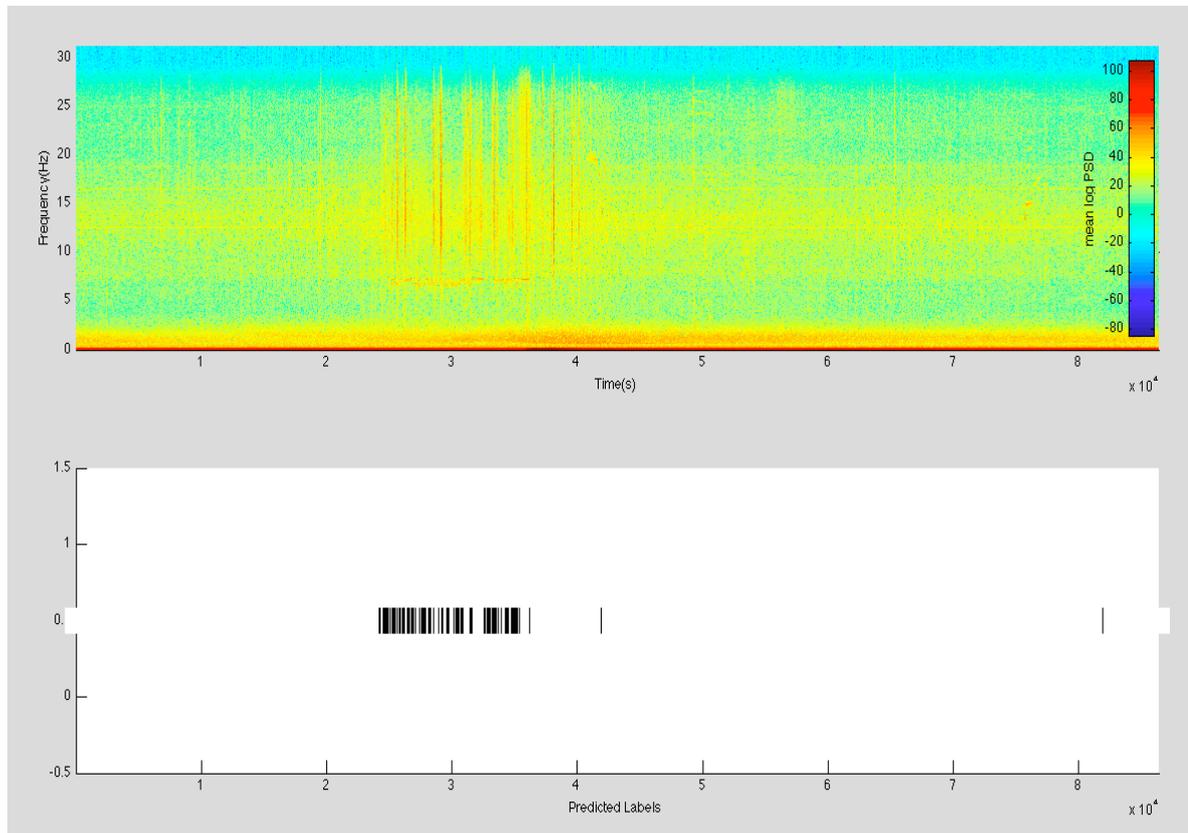

**Figure 13 : Spectrogram of the seismogram for the entire day (5-12-2002) and the corresponding presence of pump signals estimated by black predicted labels**

The black predicted labels indicate the presence of pump's seismic signals estimated by SVM1 at the corresponding time interval as shown in Figure 13. Incidentally, the predicted labels coincide with the original pump's duration of activity, thereby validating the SVM model and the application of this machine on such problems to some extent.

### 3.4.2 Testing of data with diverse interruption

The seismogram (sampling rate = 62.5 Hz) for the entire day was recorded on 7$^{th}$ December, 2002 and it was tested by SVM1 .

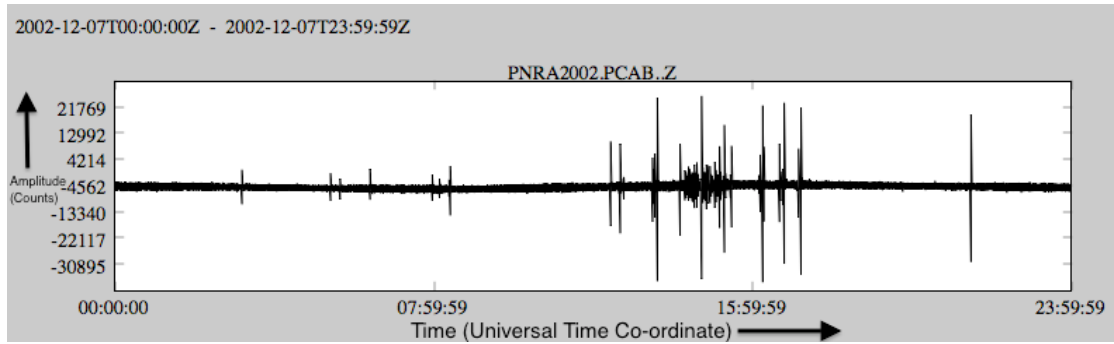

Figure 14 : Seismogram of 7-12-2002 with diverse interruption by pump signals

The machine showed that 97.3637% of the signals do not contain pump's seismic signals. It was known that the pump had been active during the day but the exact time interval of it's activity was not known which has been estimated by the Support Vector Machine by the black predicted labels.

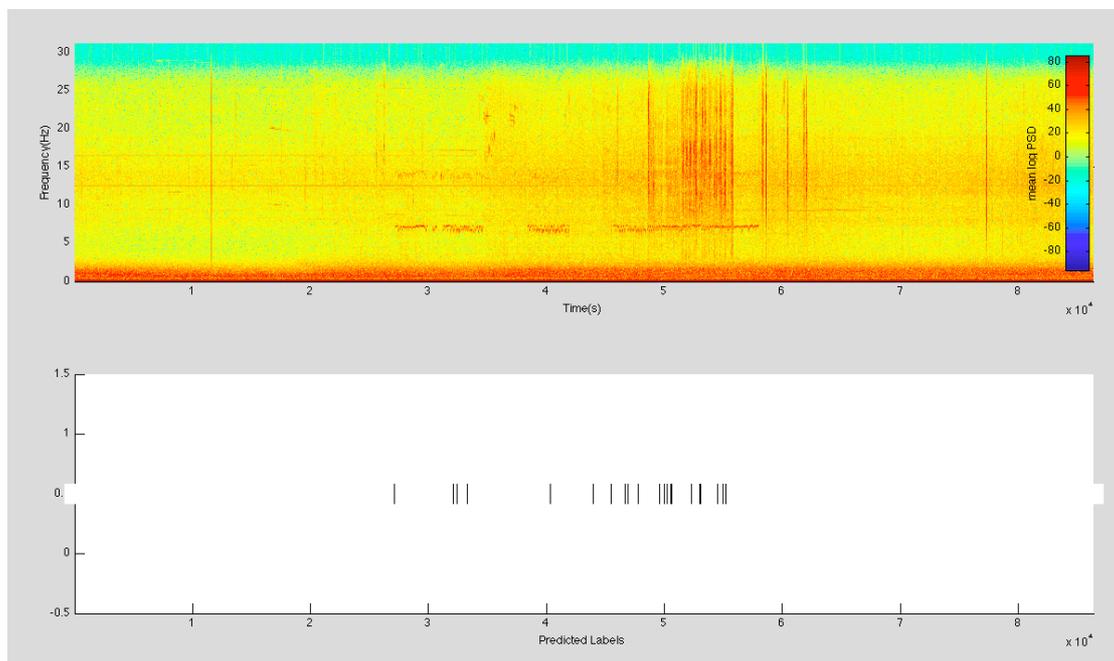

Figure 15 : Spectrogram of 7-12-2002 with corresponding predicted labels estimated by SVM1

### 3.4.3 Testing of data used in Nicola Alessandro Pino et al, 2011.

The seismogram (sampling frequency = 31.25 Hz) recorded by PCAB station on 4$^{th}$ April , 2003 has been used in Nicola Alessandro Pino et al , 2011 to prove the existence of deep accumulation of gas and slug upraise in the Stromboli area from a paroxysmal explosion's seismic precursors which was basaltic in nature. When the data was tested by SVM2, the machine showed complete absence of pump's seismic signals thereby ,asserting and further authenticating their work.

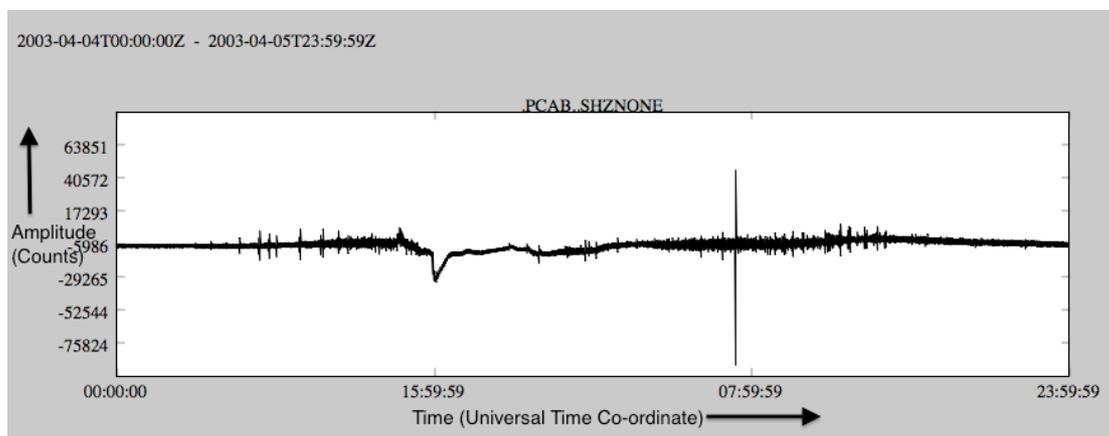

**Figure 16: Seismogram used by Nicola Alessandro Pino et al, 2011**

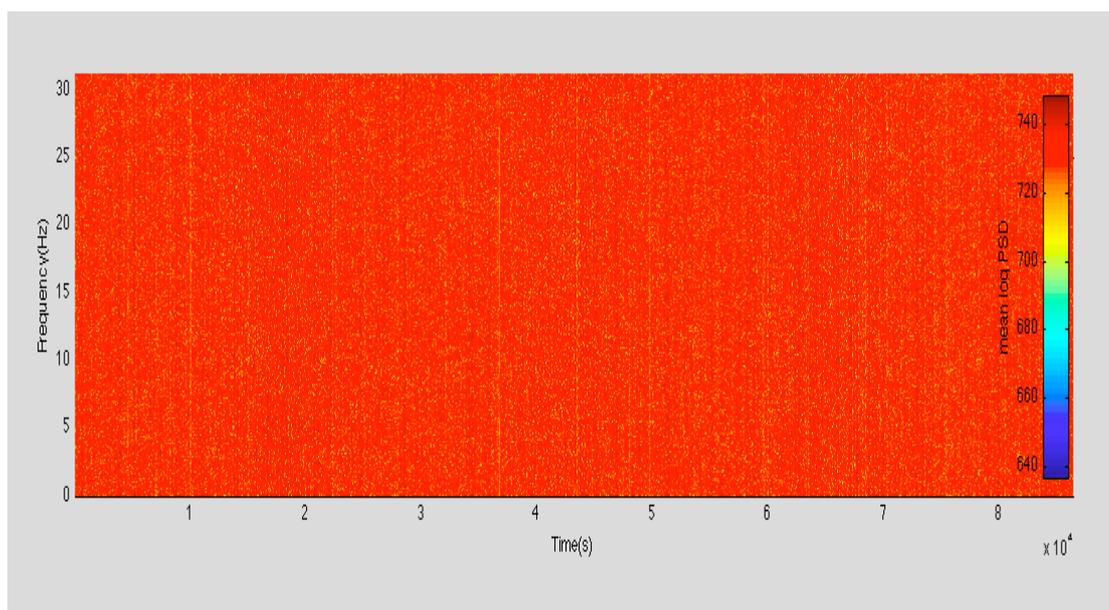

**Figure 17: Spectrogram of 04-04-2003**

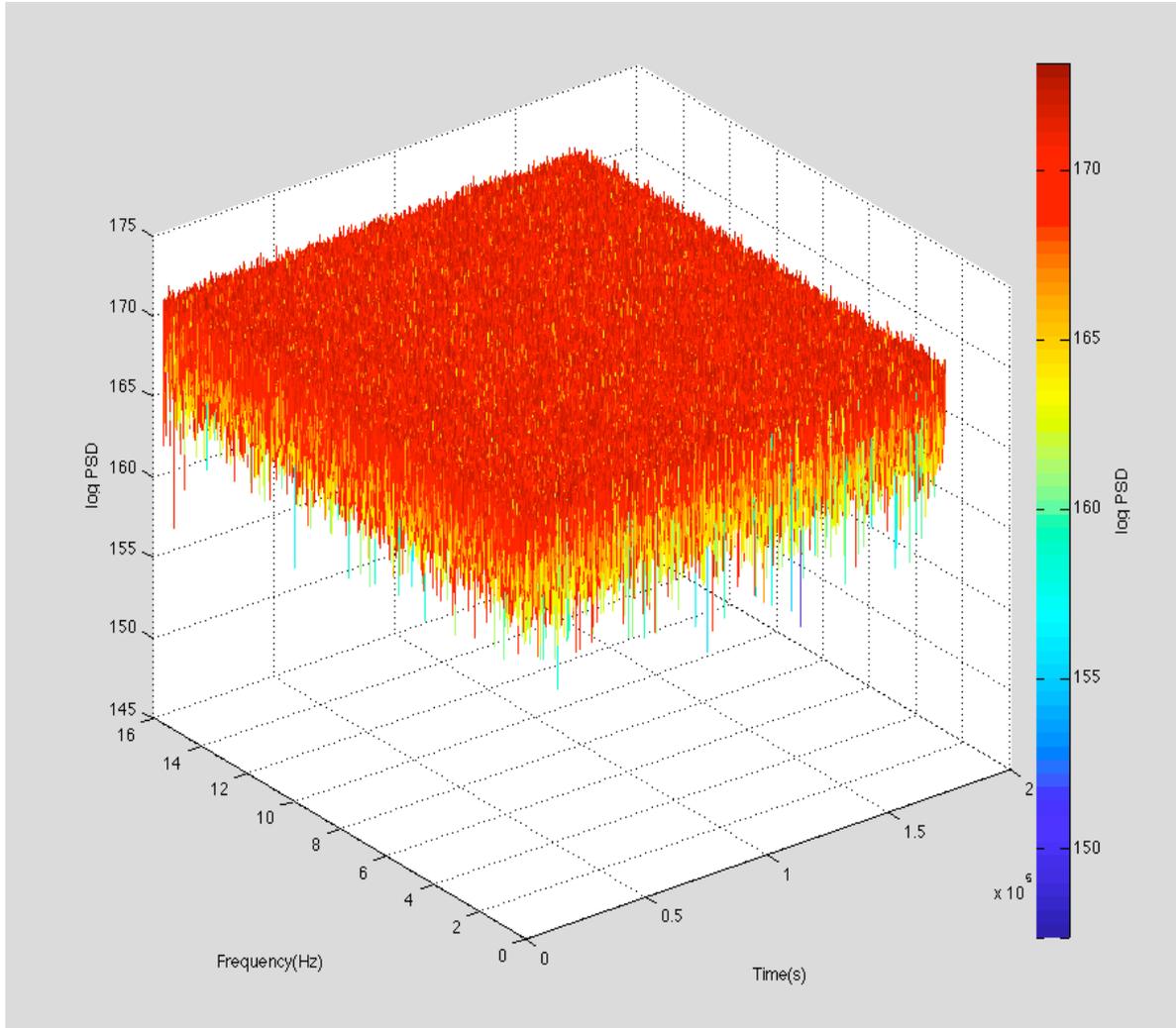

**Figure 18: 3-dimensional spectrogram of 04-04-2003**

# 5. Discussion

The deterministic approach taken by us by using Support Vector Machines has overall shown positive results. Apart results from the machine showing 99.7149% accuracy (determined through cross validation) on the parent class, which was used to train the model, the machine when tested the entire time series data of the spectrogram obtained from the seismogram of the entire day, it showed that 92.557% of the seismic signals didn't contain the pump's seismic signals as well as it's prediction of the pump's seismic signals presence (which can be seen in figure 13 marked by black labels) coincides with the observed duration of activity of the pump. This shows that the machine trained by us is authentic and is ready for further testing. On 7$^{th}$ December, 2002 it was again reported the pump was switched on but the duration of activity was unknown. On using SVM by testing the spectrogram (Power spectral density) of the seismogram for the entire data we found that 7$^{th}$ December, 2002's seismic signals had 97.3637% absent pump's seismic signals. Although, the black predicted labels in figure 15 (giving signs of pumps) are spread throughout the day indicating occurrence of diverse interruption by pump's seismic signals which had also been reported by Nicola Aleesandro Pino before the research work started.

The seismogram of 4$^{th}$ April, 2003 recorded by the same PCAB station (whose recordings have been used by us for training and testing) had been used by Nicola Alessandro Pino et al in 2011 where the seismic precursors had been used by them for suggesting the presence of slug uprise and accumulation of gas deep in the Earth. The authors were of the view that pump's seismic signals might have corrupted the seismogram, hence, rendering the research work done by them doubtful. The seismogram recorded on that day had a 31.25 Hz as its sampling frequency which is exactly half of 62.5 Hz, the sampling frequency of the seismogram used for training the Support Vector Machine. For testing 4$^{th}$ April's seismogram a new machine was trained (SVM2) by reducing the samples of the seismogram matrix by half. The algorithm used was :

$$S_i = (s_{2i+1} + s_{2i+2})/2$$

Where $i \in \{0,1,2,3,4,\ldots\ldots N\}$

Here N being the number of samples in the seismogram's matrix for pump's seismogram and no pump's seismogram, reduced separately and then used for training after calculating power spectral density.

SVM2 showed an accuracy of 96.4612 % during cross-validation. SVM2 tested the 4th April's spectrogram (power spectral density) and found that the signals didn't contain the pump's seismic signals at all thereby subverting the ambiguity and clarifying Nicola Alessandro Pino et al's work.

# 6. Conclusions

The result shows 99.7149% accuracy through cross validation after optimizing using grid search which is quite high for a supervised classification technique. SVM can have it's utility in online data processing too. On the other hand past data records can also be tested using SVM. SVM can become a useful tool in classification between signals, which contain noise and signals which do not contain noise in various fields of Geosciences which involve signal processing. One can supervise the learning machine by doing a priori labeling of patterns containing noise and then train the machine using noise containing and noise-free signals. Overall, the performance of SVM has been interestingly Sui Generis.

# Appendix

Commands used in command shell of Obspy for extracting seismogram matrix (matlab format) from raw data :

In [1]: from obspy.core import read

In [2]: from scipy.io import savemat

In [3]: from obspy.core import UTCDateTime

In [4]: from obspy.core import UTCDateTime

In [5]: st = read("/Users/apple/Desktop/PANAREA/Panarea021205/pcab.021205.z")

In [6]: dt = UTCDateTime("20021205T0655")

In [7]: st = read("/Users/apple/Desktop/PANAREA/Panarea021205/pcab.021205.z", starttime = dt , endtime = dt+2*60*60+7*60)

In [8]: for i, tr in enumerate(st):
    mdict = dict([[j, str(k)] for j, k in tr.stats.iteritems()])
    mdict['data'] = tr.data
    savemat("data-%d.mat" % i, mdict)

The above commands generate a seismogram matrix of matlab format which can be read in matlab where it can be plotted and further analyzed by usage of various mathematical functions.

So, we generate two seismogram matrices using UTCDateTime method/function for slicing the raw data from 06:55 to 09:02 (as we know that the pump signals are present in the raw data during this time interval) and from 12:00 to 14:07 . One is having pump signals (pumpo.mat) and the other is not having pump signals (nopumpo.mat) .

**In Matlab :**

**For normalization :**

```
D = xdata;

[m,n] = size(D);

Su = sum(D);

for i = 1:m , for j = 1:n , N(i,j) = (D(i,j)/Su(1,j));
   end
end

N = N';
```

**For generating training data :**

```
>>load pump;
x = double(data);
[S,F,T,P] = spectrogram(x,1024,512,1024,62.5);
v = P(3:202,:)';

load nopump;
x = double(data);
[S,F,T,P] = spectrogram(x,1024,512,1024,62.5);
nv = P(3:202,:)';
xdata = vertcat(v,nv);
```

```matlab
D = xdata;

[m,n] = size(D);

Su = sum(D);

for i = 1:m , for j = 1:n , N(i,j) = (D(i,j)/Su(1,j));
   end
end

N = N';

tdata = sparse(N);
for j=1:929, c(j)=1;
end
c = c';
for j=1:929, nc(j)=-1;
end
nc = nc';
group = vertcat(c,nc);

save('train','tdata','group');
```

---

The above program generates a file which can be used in svm program for carrying out training .

**For generating testing data in matlab :**

```matlab
load data;
x = double(data);
[S,F,T,P] = spectrogram(x,1024,512,1024,62.5);
```

```matlab
test = P(3:202,:)';
D = xdata;

[m,n] = size(D);

Su = sum(D);

for i = 1:m , for j = 1:n , N1(i,j) = (D(i,j)/Su(1,j));
    end
end

N1 = N1';

test = sparse(N1);
for j=1:10545 , group1(j) = -1;
end
group1 = group1';

save('test','test','group1');
```

---

The above program generates a file which can be used in svm program for carrying out testing. Here data.m is the seismogram matrix which has to be classified by svm classifier generated by ObsPy.

The svm program used is a unix based program from libsvm-3.16 package which can be downloaded from :

http://www.csie.ntu.edu.tw/~cjlin/libsvm/

and for running the "makefile" one needs a gcc compiler for linux platforms, x code for macintosh platforms, for windows the program is already available in the package and other c++, matlab, python plug-ins are also available within the package. After running the Makefile, "svm-train", "svm-predict", "svm-scale" are automatically created.

**Training :**

In unix shell :

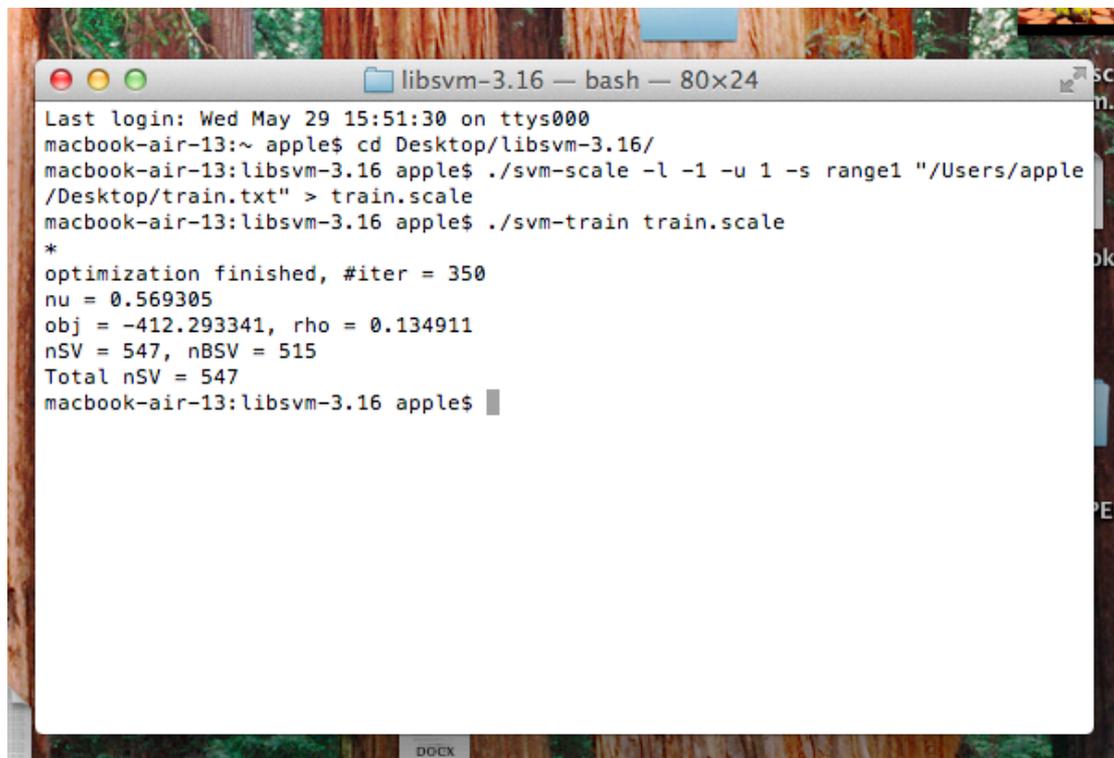

macbook-air-13:~ apple$ cd Desktop/libsvm-3.16/

macbook-air-13:libsvm-3.16 apple$ ./svm-scale -l -1 -u 1 -s range1 "/Users/apple/Desktop/train.txt" > train.scale

macbook-air-13:libsvm-3.16 apple$ ./svm-train train.scale

*

optimization finished, #iter = 350

nu = 0.569305

obj = -412.293341, rho = 0.134911

nSV = 547, nBSV = 515

Total nSV = 547

First the data is scaled and then it is trained , after this process a "train.scale.model" file is created which can be used for testing purposes

**Testing :**

macbook-air-13:libsvm-3.16 apple$./svm-scale -r range1 "/Users/apple/Desktop/test.txt" > test.scale

macbook-air-13:libsvm-3.16 apple$ ./svm-predict test.scale train.scale.model test.predict

Accuracy = 90.625% (841/928) (classification)

macbook-air-13:libsvm-3.16 apple$

One can generate predicted labels by adding :

```
printf("%g\n",predict_label);
```

in 138th line of svm-predict.c file of libsvm-3.16 , obtained from libsvm package and then compiling it again to generate the svm-predict program.

**For generating seismogram and then integrating it with predicted labels pictorially. (in Matlab) :**

```
for i = 1:10545 , yn(1,i) = 0.5;
end

for i = 1:10545 , if numneric(i,1)>0 numneric(i,1)= 90;
   else numneric(i,1) = 30;
   end
end;
```

```
zn = numneric';
x = double(data);
figure;
h(1) = subplot(2,1,1);
[y,f,t,p] = spectrogram(x,1024,512,1024,31.25,'yaxis');
surf(t,f,10*log10(abs(p)),'EdgeColor','none');
axis xy; axis tight; colormap(jet); view(0,90);

h(2) = subplot(2,1,2);
dotsize = 25;
scatter(t(:),yn(:), dotsize, zn(:));
linkaxes(h,'x');
```